\documentclass[12pt]{iopart}
\usepackage{times}
\usepackage{graphicx}

\begin{document}

\title{Shot noise in generic quantum dots:
crossover from ballistic to diffusive transport}
\author{H.-S. Sim and H. Schomerus}
\address{Max-Planck-Institut f\"{u}r Physik komplexer Systeme,
N\"{o}thnitzer Str. 38, 01187 Dresden, Germany}
\begin{abstract}
We study shot noise for
generic quantum dots coupled to two leads
and allow for an arbitrary strength of diffractive impurity
scattering inside the dots.
The ballistic quantum dots possess a mixed classical phase space,
where regular and chaotic regions coexist. 
In absence of disorder,
the noise is systematically suppressed below the
universal value of fully chaotic systems, 
by an amount which varies with the positions of the leads.
This suppression is due to the deterministic nature of transport 
through regular regions and along short chaotic trajectories,
and disappears by increasing the scattering rate,
which is incorporated by using the Poisson kernel of
random matrix theory.
\end{abstract}

\section{Introduction}
\label{INTRO}
The discreteness of the electron charge $e$ causes
time-dependent current fluctuations even at zero temperature.
These fluctuations are known as shot noise and
recently have received much attention
in mesoscopic physics
(for a review see \cite{Blanter_Phyrep}).
One prominent class of mesoscopic systems are
classically chaotic ballistic cavities (quantum dots).
For incoherent transport through a chaotic quantum dot 
coupled to two leads, each having $N$ channels,
the shot noise would assume the
Poissonian value $P_0 =2 e G_0 V$, where $G_0=N e^2/ (2 h)$ and
$V$ is the voltage difference between the leads.
For low temperatures,
correlations of electrons due to Fermi statistics 
suppress the noise $P$ by a factor of ${\cal F}=P/P_0$
relative to the value $P_0$ of uncorrelated electrons.
In chaotic quantum dots,
the suppression factor ${\cal F}={\cal F}_{\rm ch}=1/4$
has been found to be
universal, i.e., independent of the details of systems, 
by using random matrix theory \cite{Jalabert,Beenakker_RMP}
and semiclassical methods \cite{Blanter_PRL},
which has been confirmed by an experiment \cite{Oberholzer}.
The origin of the low-temperature noise in chaotic quantum dots
is the {\em probabilistic} nature of quantum transport:
Each attempt to transmit a charge $e$ through the system
succeeds with a finite probability.

Very recently, further reductions of shot noise below 
${\cal F}_{\rm ch}$ due to residual signatures of classically
{\em deterministic} scattering 
have been discussed by Agam, Aleiner, and Larkin \cite{Agam}.
This non-universal suppression of the noise
appears when
the dwell time of electrons is smaller than the 
time scale of quantum diffraction 
(the Ehrenfest time in chaotic systems).
This has also been verified in a recent experiment \cite{Oberholzer2}.

Generic ballistic quantum dots are not fully chaotic but
possess a mixed classical phase space \cite{Markus}, 
where regular islands
are separated from chaotic seas by impenetrable dynamical barriers.
Signatures of the mixed phase space in quantum transport
have been found in the conductance,
which exhibits fractal fluctuations \cite{Ketzmerick} 
and isolated resonances \cite{Huckestein,Hufnagel}. 
In this proceeding paper, we address 
the effect of the mixed phase space and of disorder
on the shot-noise suppression.

We study shot noise for generic quantum dots coupled to two leads,
with varying the positions and the widths of the leads,
and allow for an arbitrary strength of elastic diffractive impurity
scattering inside the dots.
In Ref. \cite{Sim} this combination of effects was used to
investigate the time scales of deterministic transport
through the quantum dots.
In the present paper we concentrate on the effect of impurity scattering
itself.
In absence of impurity scattering,
the suppression factor ${\cal F}$ is systematically reduced below the
universal value ${\cal F}_{\rm ch}$ of fully chaotic systems, 
by an amount which varies with the parameters
(positions and widths) of the attached leads.
This reduction of ${\cal F}$ is
not only due to the deterministic nature of transport
along short chaotic trajectories,
but also because quantum diffraction
is strongly reduced for transport through regular regions, as well,
thus, the dependence of ${\cal F}$ on the parameters of leads
can be understood from the fact that the amount of deterministic 
processes coupled to the leads varies with the parameters.
This feature becomes clarified after adding
diffractive impurity scattering with the aid of
the Poisson kernel
\cite{Beenakker_RMP,Mello,Brouwer,Baranger_Poisson}
of random matrix theory.
As the rate of diffractive scattering increases,
the suppression factor ${\cal F}$ becomes larger
and eventually recovers the universal value
${\cal F}_{\rm ch}$ of fully chaotic systems.

In our numerical simulation, the annular billiard
\cite{Bohigas} is chosen as
a model of a ballistic quantum cavity with mixed phase space.
We will discuss the classical motion in the billiard 
in Sec. \ref{ANNULAR}, and
incorporate diffractive scattering in Sec. \ref{DYNAMIC}.
In Sec. \ref{RESULT},
the dependence of ${\cal F}$ on the parameters of the leads
will be discussed in absence and presence of diffractive scatterings,
and
the role of regular versus chaotic classical motion on
shot noise will be clarified.
In Sec. \ref{CONCLUSION}, finally, some concluding remarks will be
given.

\section{Annular billiards}
\label{ANNULAR}

An exemplary model for mixed regular and chaotic
classical dynamics is the two-dimensional annular billiard
\cite{Bohigas},
which consists of 
the region between two
circles with radii $R$, $r$, and eccentricity $\delta$.
Its schematic diagram is drawn in Fig.\ \ref{fig1}(a).
To study shot noise,
two leads (openings) of width $W$
are attached opposite to each other at an angle 
$\theta \in [0,\pi/2]$ with respect to
the axis through the two circle centers.
Two values of $W$ are selected as
$W=0.222 R$ and $W=0.347 R$;
the former corresponds to leads with six channels,
while the latter to ten channels.
The other parameters of $r$ and $\delta$ are fixed as
$r=0.6 R$ and $\delta=0.22 R$, respectively,
in terms of $R$ throughout this work.

\subsection{Classical motion in annular billiards}

\begin{figure}
\begin{center}
\includegraphics[width=0.80\textwidth]{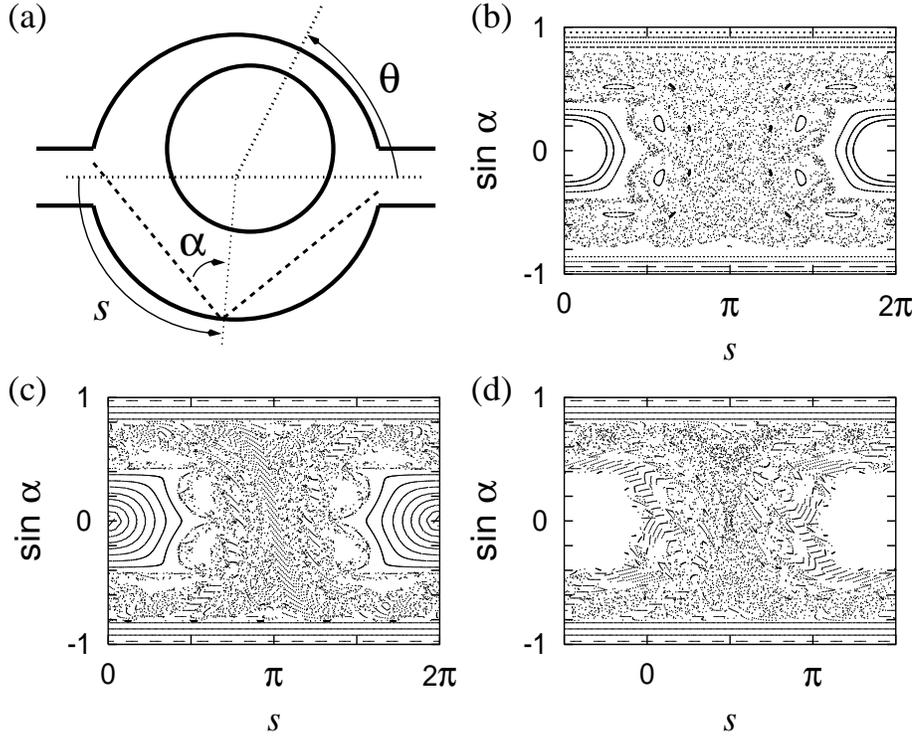}
\end{center}
\caption{
(a) Schematic diagram of the annular billiard between
two circular hard walls of exterior radius $R$, interior radius
$r=0.6 R$, and eccentricity
$\delta = 0.22 R$. Two openings of width $W=0.222 R$ are attached 
opposite to each other at an
angle $\theta \in [0,\pi/2]$
relative to the line connecting the circle centers.
(b) Phase space, parameterized by the impact
parameter $s$ and the transverse momentum component $\sin\alpha$
of trajectories at the exterior circle.
The lower panels show the
phase space for open billiards with $\theta=0$ (c) and $\theta=\pi/2$
(d), for which we only record trajectories that are injected through the
openings
(until they leave the billiard again).
}
\label{fig1}
\end{figure}

The classical phase space of the annular billiard can be parameterized by
the impact parameter $s$ and the transverse component
of the momentum $\sin\alpha$ (with $\alpha$
the angle of incidence) of trajectories that are reflected 
at the exterior circle. 
The phase space of the closed annular billiard 
is shown in Fig.\ \ref{fig1}(b), which 
displays two regular 
whispering-gallery regions, a large regular island, 
neighboring satellite islands, and a chaotic sea.
Figs.\ \ref{fig1}(c) and (d) show the phase space of the open annular
billiard (with opening width $W=0.222 R$),
which only includes trajectories of particles
that are injected into the billiard
through leads attached at $\theta=0$ and $\theta=\pi/2$, respectively.
The large island is well coupled to 
one of the openings for $\theta=0$,
and is completely decoupled from both openings at $\theta=0.5\pi$
[see the empty region at the center of Fig.\ \ref{fig1}(d)].
The chaotic sea and whispering-gallery regions are well coupled 
for arbitrary position of the openings,
while the coupling of the satellite islands depends on $\theta$ and $W$.
In this way one can select regions in phase space
by varying $\theta$ and $W$.

In the open billiard
some classical trajectories entering into the chaotic sea 
escape without many billiard-boundary scatterings
so that
the chaotic sea appears less chaotic than
in the closed billiard.
One can see this modification of phase space,
e.g., near $s=\pi$ in Fig.\ \ref{fig1}(c) 
and around the empty center in (d).
The role of these short chaotic trajectories on shot noise will be
discussed later.

\subsection{Conductance and shot noise formula}
\label{ANNULAR2}

We calculated the dimensionless conductance $T$ and
the shot-noise suppression factor ${\cal F}$
from the relations \cite{Blanter_Phyrep}, 
\begin{eqnarray}
T & = & {\rm tr}\, t^\dagger t, \\
{\cal F} & = & (2/N)\,{\rm tr}\, t^\dagger t(1-t^\dagger t).
\label{noise_eq}
\end{eqnarray}
Here, $N$ is the number of channels in each lead and
the transmission matrix $t$ 
is a subblock of the scattering matrix
\begin{equation}
S=\left(\begin{array}{cc}r&t'\\t&r'\end{array}\right),
\end{equation}
where $r$ and $r'$ are reflection coefficients and the transmission
matrix $t'$ contains the same information as $t$.
The scattering matrix is obtained numerically by the method 
of recursive Green functions \cite{Baranger_method}, for which space is 
discretized on a square lattice. In terms of the lattice constant
$a$, we choose
$R=144\,a$, $r=86.4\,a$, and $\delta=31.7\,a$.
Energy $E$ will be measured in units of $\hbar^2/(2 m a^2)$
and time in units of $2 m a^2/\hbar$, with $m$
the mass of the charge carriers. In these units
the mean level spacing $\Delta$ is estimated as 
$\Delta = 4\pi / A \simeq 0.0003$,
where $A$ [$= \pi(R^2-r^2)/a^2$] is the dimensionless area of billiard.
We will work in the energy window  $E\in(0.408,0.433)$.
In this window, the Fermi wavelength $\lambda_{\rm F}\simeq 9.5\,a$,
resulting in
$N= {\rm Int}(2W/\lambda_{\rm F})=6$ for $W=0.222 R=32\,a$ and
$N= 10$ for $W=0.347 R=50\,a$.


\section{Elastic diffractive scatterings}
\label{DYNAMIC}

We now introduce elastic diffractive scatterings
with the aid of
the Poisson kernel \cite{Mello,Brouwer,Baranger_Poisson},
a statistical ensemble of random
matrix theory \cite{Beenakker_RMP}.
In the Poisson kernel one averages the scattering matrix over an 
energy range $E_{\rm av}$, in this way eliminating 
the system-specific details of the dynamics 
with
time scales longer than $t_{\rm av}=\hbar / E_{\rm av}$,
and replaces these by random dynamics of the same universality class
as elastic diffractive impurity scattering.
The effective mean free scattering
time $t_{\rm av}$ can be tuned by changing
the energy-averaging window $E_{\rm av}$.

In our numerical simulation,
the elimination of the system-specific details on time scales
larger than $t_{\rm av}$
is achieved by averaging the scattering matrix $S$ over an energy range
$[E_0 - E_{\rm av}/2,E_0 + E_{\rm av}/2]$ of width 
$E_{\rm av} =\hbar/t_{\rm av}$ 
(which will be taken inside the total energy  range [0.408,0.433]
of our numerical simulation),
\begin{equation}
\overline{S}(E_{\rm av};E_0)= E_{\rm av}^{-1}
\int_{E_0 - E_{\rm av}/2}^{E_0 + E_{\rm av}/2}
 dE\,S(E).
\end{equation}
Here the information on processes with longer time scales
than $t_{\rm av}$ is lost, because it is encoded
into the short-range energy correlations (fluctuations) 
of the scattering matrix, 
while the information on the dynamics on
shorter time scales than $t_{\rm av}$ modulates the scattering matrix on
larger energy scales and hence is retained. 
Diffusive motion then is
introduced
by
coupling to an auxiliary system
with scattering matrix
$S_0$ taken from the appropriate circular ensemble
of random matrix theory (observing
the same symmetries as the original scattering matrix, as time-reversal
or spatial parities \cite{Baranger_Sym}), resulting in
\begin{eqnarray}
S'(E_{\rm av};E_0;S_0) = \overline{S}(E_{\rm av};E_0) + 
{\cal T}^\prime (1-S_0{\cal R})^{-1}S_0{\cal T}.
\label{smatrix}
\end{eqnarray}
The ensemble of scattering matrices (\ref{smatrix}) is the so-called
Poisson kernel \cite{Beenakker_RMP,Mello,Brouwer,Baranger_Poisson},
with $\overline S$ the so-called optical scattering matrix.
The coupling matrices ${\cal T}$, ${\cal T}^\prime$, and ${\cal R}$
must be chosen such that $S^\prime$ is a unitary matrix,
but the
invariance of the circular ensemble guarantees that
results do not depend on their specific choice.

Then, the mean suppression factor ${\cal F}(E_{\rm av})$ 
(or the conductance $T$) can be obtained
for fixed $E_{\rm av}$ (and hence fixed $t_{\rm av}$), first
by averaging the noise (or the conductance)
within each Poisson kernel (fixing also $E_0$),
and then averaging these values over $E_0\in (0.408,0.433)$.

\begin{figure}
\begin{center}
\includegraphics[width=0.80\textwidth]{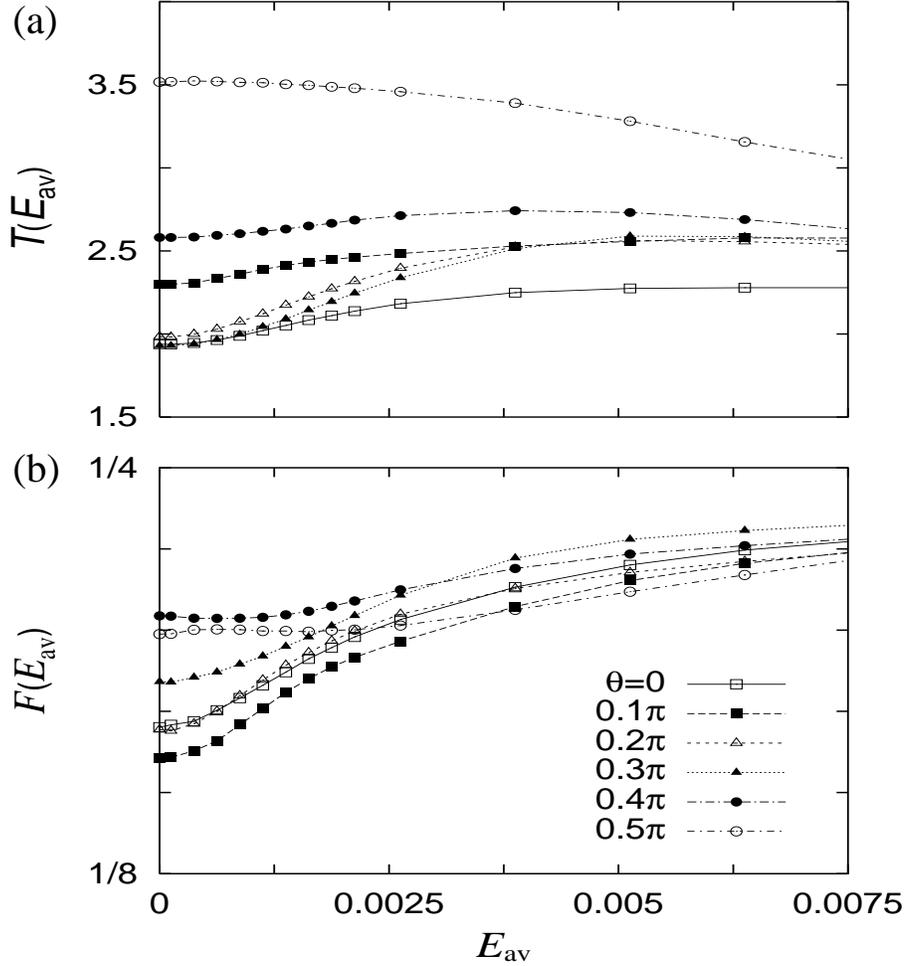}
\end{center}
\caption{
(a) Transmission $T$ and (b) shot-noise suppression factor
${\cal F}$ (lower panel) as a function of the strength of diffractive
impurity scattering with a rate $\simeq E_{\rm av}/\hbar$,
for different positions $\theta$ of the leads with $W=0.222R$ ($N=6$). 
In (a), the same symbols are used for each $\theta$ as in (b).
}
\label{fig3}
\end{figure}

\section{Results and discussions}
\label{RESULT}

\subsection{Absence of diffractive scatterings:
the dependence of shot noise on the positions and on the widths
of the leads}
\label{ABSENCE}

\begin{figure}
\begin{center}
\includegraphics[width=0.80\textwidth]{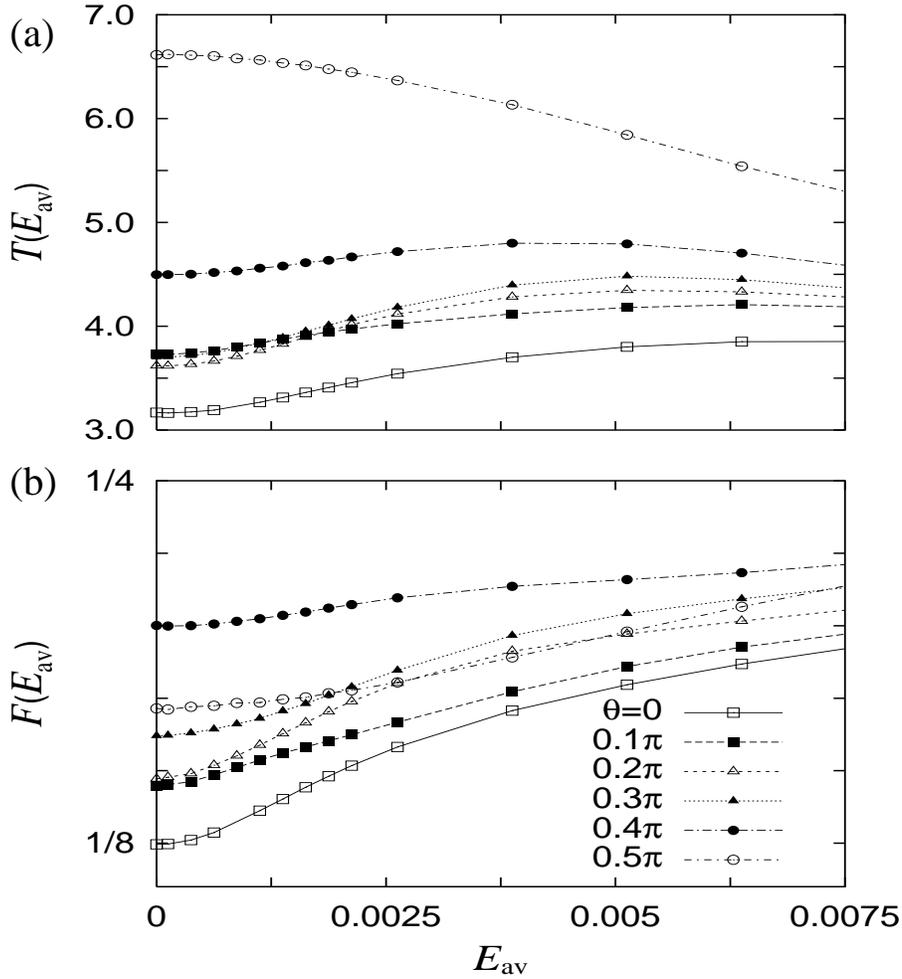}
\end{center}
\caption{
The same as Fig. \ref{fig3}, but for $W=0.347R$ ($N=10$).
}
\label{fig4}
\end{figure}

The energy-averaged conductance $T$ and
the energy-averaged suppression factor ${\cal F}$ 
in the window of $E\in(0.408,0.433)$
are shown as the values at $E_{\rm av}=0$ 
of Figs. \ref{fig3} and \ref{fig4}
for different $\theta$ and $W$.
Our first observation for the shot noise is that
for all values of $\theta$, the suppression factor is smaller than the
universal value $1/4$ of fully chaotic motion.
Moreover, the
shot noise is more suppressed for the cases that one opening couples to 
the large regular island ($0 < \theta \leq 0.3\pi$) than for the cases
in which the
regular islands is decoupled from the openings ($\theta \geq 0.4\pi$).
This behavior indicates that electrons injected into the large regular 
island contribute less to the shot noise, which will be
clarified in the analysis incorporating
diffractive scatterings in Sec. \ref{PRESENCE}.

The suppression factors are strongly reduced
for $W=0.347R$ than for $W=0.222R$,
as shown in Figs. \ref{fig3} and \ref{fig4}.
This feature will be also understood 
in Sec. \ref{PRESENCE}
from the fact that
the amount (per channel in the leads) of deterministic trajectories,
which include those in the regular islands 
and those escaping the dots without many billiard-boundary scatterings
in the chaotic sea,
increases for the leads with broader widths.

We note that the dependence of conductance $T$ on $\theta$ shows
a similar features to that of the suppression factor;
conductance is smaller for the well-coupled cases to the regular
islands ($0 < \theta \leq 0.3\pi$) than for the decoupled cases
($\theta \geq 0.4\pi$).
This behavior can be understood from the fact that
for $0 < \theta \leq 0.3\pi$ the regular
island is well coupled only to one opening, so that
electrons entering the regular island do not contribute to 
transmission.
This similarity, however, is accidental.
As will be demonstrated in Sec. \ref{PRESENCE},
deterministic processes will always suppress shot noise
(when their area in phase space is larger than a Planck cell, 
so that they are well resolved),
while they enhance or suppress conductance, depending on the
shape of the quantum dot and the positions of the leads.

\subsection{Role of regular versus chaotic classical motion
in shot-noise suppression}
\label{PRESENCE}

The effect of the diffractive scattering on the conductance $T$ 
and the shot-noise suppression factor ${\cal F}$
for different positions of the
leads are shown in Figs.\ \ref{fig3} and \ref{fig4}.

The values of $T$ and ${\cal F}$ at $E_{\rm av}=0$ are those in
absence of diffactive impurity scattering,
because the Poisson kernel
becomes a delta function at $S(E_0)$ as $\overline S$ approaches 
this unitary matrix for $E_{\rm av}\to 0$.
For increasing $E_{\rm av}$, $T(E_{\rm av})$ and ${\cal F}(E_{\rm av})$
approach their universal values of random-matrix
theory, because $\overline S=0$ in this limit.
Note that spatial symmetries such as up-down and left-right reflections
affect the weak localization corrections to the universal values
\cite{Baranger_Sym}.
The correction is enhanced (absent) for $\theta=0$ ($0.5\pi$), 
where there is up-down (left-right) reflection
symmetry \cite{Baranger_Sym},
compared with the other values of $\theta$ with no spatial symmetry.

The function ${\cal F}(E_{\rm av})$ is monotonically increasing
almost everywhere (negative values of the slope are within the
statistical uncertainty in the numerical simulation),
indicating 
that the noise is {\it enhanced} when the
system-specific deterministic details of the transport
are replaced by the
indeterministic transport of random matrix theory.
Consequentially, one can define the probability distribution
of deterministic processes with dwell time $t_{\rm dwell} \simeq t_{\rm av}$
by the rate of change of the shot-noise suppression
${\cal F}(E_{\rm av})$
which can be substantiated by a semiclassical analysis
\cite{Sim}.

The crossover of shot noise from ballistic to diffusive transport
occurs at different scattering rate $\hbar/E_{\rm av}$,
depending on $\theta$ and $W$.
We will discuss first for the case of $W=0.222R$.
In the three cases $\theta=0$,
$0.1\, \pi$,
$0.2\,\pi$, in which the leads couple to the large regular island in
phase space, 
the crossover occurs already at very small $E_{\rm av}$,
while it starts above an energy $E_{\rm indet} \approx 0.002$ for
$\theta=0.4\,\pi$, $0.5\,\pi$ (where the regular region is decoupled).
The former is due to the deterministic processes of very long dwell times 
in the large regular island, which can be affected by diffractive
scatterings of very small scattering rate.
On the other hand, the latter is due to the fact that
the processes in the chaotic sea
are already
indeterministic for times larger than $\hbar/E_{\rm indet}$,
which can be interpreted as the Ehrenfest time.

For the leads with larger width $W$, 
more deterministic processes of regular regions can
couple to the leads and
more short deterministic trajectories in the chaotic sea
appear.
This feature entails that
for the leads with $W=0.347R$ (Fig. \ref{fig4}), 
${\cal F}(E_{\rm av})$ approaches
to ${\cal F}_{\rm ch}$ more slowly than for the case of
$W=0.222R$ (Fig. \ref{fig3}).
Similarily,
for the leads at $\theta=0.5 \pi$ and with $W=0.347R$
the suppression factor ${\cal F}(E_{\rm av})$
starts to increases at earlier $E_{\rm av}$ than
$E_{\rm indet}$ (which is obtained in the curves for $W=0.222R$),
since small regular islands additionally couple to the leads with
$W=0.347R$.

We note that
in contrast to ${\cal F}(E_{\rm av})$, the sign of the slope of 
$T(E_{\rm av})$ depends on the system-specific geometry, e.g., 
on lead position $\theta$ [see Figs.\ \ref{fig3}(a) and \ref{fig4}(a)]. 
This shows that conductance can be either enhanced or suppressed
by diffractive scattering (indeterministic processes).

%

\section{Concluding remarks}
\label{CONCLUSION}

The shot noise reduction due to
deterministic transport suggests that
shot noise can be
used as a probe of classical phase space of generic quantum dots
(conductance can not be used as such a probe).
The crossover of shot noise from ballistic to
diffusive transport, shown in
Figs.\ \ref{fig3}(b) and \ref{fig4}(b),
could be probed experimentally by
measuring shot noise while tuning the disorder strength,
which can be achieved by adjusting a gate voltage \cite{toyoda}.

\section*{References}

\end{document}